\documentclass[12pt]{article}
\usepackage[utf8]{inputenc}
\usepackage{amsfonts}
\usepackage{amsmath}
\usepackage{amssymb}
\usepackage{graphicx}
\usepackage{cite}
\usepackage{color}
\usepackage[left=1cm,top=1cm,bottom=1cm,right=1cm,nohead,nofoot]{geometry}

\def \be {\begin{equation}}
\def \ee {\end{equation}}
\def \bea {\begin{eqnarray}}
\def \eea {\end{eqnarray}}
\def \nn {\nonumber}

\def \rr {\raise.35ex\hbox{\small $\prime$}\kern-.17em{\mbox{\large $\imath$}}}

\def \dels {\partial\kern-.6em /\kern.1em}
\def \As {{A\kern-.5em / \kern.5em}}
\def \Ds {D\kern-.7em / \kern.5em}

\def \ks {k\kern-.5em /}
\def \ls {l\kern-.5em /}

\reversemarginpar
\newcommand{\hide}[1]{}

\begin{document}
\begin{titlepage}

\begin{center}

\hfill
\vskip .2in

\textbf{\LARGE
Gauge Transformation of Double Field Theory for Open String
\vskip.5cm
}

\vskip .5in
{\large
Chen-Te Ma\footnote{e-mail address: yefgst@gmail.com}\\
\vskip 3mm
}
{\sl
${}$
Department of Physics, Center for Theoretical Sciences and
Center for Advanced Study in Theoretical Sciences, 
National Taiwan University, Taipei 10617, Taiwan,
R.O.C.}\\
\vskip 3mm
\vspace{60pt}
\end{center}
\begin{abstract}
We combine symmetry structures of ordinary (parallel directions) and dual (transversal directions) coordinates to construct the Dirac-Born-Infeld (DBI) theory. The ordinary coordinates are associated with the Neumann boundary conditions and the dual coordinates are associated with the Dirichlet boundary conditions. Gauge fields become scalar fields by exchanging the ordinary and dual coordinates. A gauge transformation of a generalized metric is governed by the generalized Lie derivative. The gauge transformation of the massless closed string theory gives the $C$-bracket, but the gauge transformation of the open string theory gives the $F$-bracket. The $F$-bracket with the strong constraints is different from the Courant bracket by an exact one-form. This exact one-form should come from the one-form gauge field. Based on symmetry point of view, we deduce a suitable action with a non-zero $H$-flux at the low-energy level. From an equation of motion of the scalar dilaton, it defines a generalized scalar curvature. Finally, we construct a double sigma model with a boundary term and show that this model with constraints is classically equivalent to the ordinary sigma model.
\end{abstract}

\end{titlepage}

\section{Introduction}
\label{1}
The most interesting topic of the M-theory is duality. In particular, the T-duality shows the equivalence between ordinary and dual theories by exchanging radius and reciprocal of radius. This gives a bigger structure to unify different kinds of theories. The T-duality of the closed string theory \cite{Zwiebach:1992ie} exchanges momentum and winding modes. A non-trivial problem of the T-duality is a non-geometrical feature in the massless closed string theory. The generalized geometry \cite{Gualtieri:2003dx} and double field theory (DFT) \cite{Hohm:2010pp} formulate the \lq\lq stringy geometry\rq\rq  \cite{Hatsuda:2012vm} to solve this problem. For the open string theory, the T-duality exchanges the Dirichlet and Neumann boundary conditions. A low-energy effective theory of the open string is a well-known DBI theory from quantum fluctuation of the open string boundary term. The open string theory has an equivalent description between the commutative and non-commutative parameters. The D-brane and R-R fields also play an important role to promote the T-duality to U-duality \cite{Hatsuda:2012uk}. The manifest U-duality is studied in \cite{Hatsuda:2013dya}. These formulations have a drawback on the gauge symmetry which relies on the section conditions or strong constraints (removing the dual coordinates) \cite{Ma:2014ala}. 
 
The recent development of a geometrical interpretation for the brane theory is the exotic brane theory, which shows that we need a global geometry. The world-volume exotic brane theory $5_2^2$ is already constructed in \cite{Kimura:2014upa}. The interesting exotic brane shows hope to give a new structure of low-energy effective field theories \cite{Callan:1986bc, Zwiebach:1985uq} from string point of view.

The main task of this paper is to extend a geometrical interpretation of the D-brane \cite{Asakawa:2012px} to the double formulation. We obtain the $C$-bracket without considering the one-form gauge field, but the $F$-bracket appears in our studies when including the one-form gauge field. From the B-transformation, we find that the $F$-bracket implies that the open string $\it cannot$ be described by the $O(D, D)$ structure. The primary reason is that the T-duality of the open string changes dimensions, but the T-duality of the closed string does not. The difference between the $C$-bracket and $F$-bracket with the strong constraints is the exact one-form. This exact one-form breaks the $O(D, D)$ structure. Based on symmetry point of view, we construct the D-brane theory on curved background. This action is also consistent with the one-loop $\beta$ calculation \cite{Callan:1986bc}. The generalized scalar curvature can be defined from the symmetry and equation of motion of the scalar dilaton. Finally, we propose the double sigma model with a boundary term. This double sigma model with the constraints is classically equivalent to the ordinary sigma model.

The plan of this paper is to first review the gauge transformation of the double field theory for the massless closed string theory in Sec. \ref{2} and review some basics of the D-brane theory in Sec. \ref{3}. Then we construct the double field theory of the DBI theory in Sec. \ref{4}. It includes the gauge transformation, bracket, action and the discussion of the Ricci scalar. We discuss a double sigma model in Sec. \ref{5}. Finally, we conclude in Sec. \ref{6}.

\section{Review of the Gauge Transformation of the Double Field Theory for the Massless Closed String Theory}
\label{2}
We review the gauge transformation of the double field theory for the massless closed string theory in this section. At first, we introduce convenient notations for the DFT and write down the gauge transformation for the generalized metric formulation \cite{Hohm:2010pp}.

\subsection{Basics}
The double field theory is built on the double coordinates. The ordinary coordinates are associated with the momentum modes and the other coordinates (dual coordinates) are associated with the winding modes. The field components are the metric field ($g_{MN}$), antisymmetric field ($B_{MN}$) and scalar dilaton ($d$). We have two constraints
\bea
\partial_M\tilde{\partial}^M(\mbox{field})=0,
\qquad
\partial_M\tilde{\partial}^M((\mbox{field})_1(\mbox{field})_2)=0,
\nn\\
\eea
where
\bea
\partial_M=\frac{\partial}{\partial x^M},
\qquad
\tilde{\partial}^M=\frac{\partial}{\partial\tilde{x}_M}.
\eea
The index $M=0, 1\cdots, D-1$ (We indicate the non-doubled target indices from $M$ to $Z$.). 
The constraints imply
\bea
\partial_M(\mbox{field})_1\tilde{\partial}^M(\mbox{field})_2+\tilde{\partial}^M(\mbox{field})_1\partial_M(\mbox{field})_2=0.
\nn\\
\eea
We need these two constraints (strong constraints) to obtain gauge invariant action up to the cubic order. If we only consider the first constraint, this constraint is the so-called weak constraint. The reason why we need the strong constraints is
\bea
\partial_M\tilde{\partial}^M\delta(\mbox{field})\neq 0,
\eea
where $\delta$ is the gauge transformation. The above relation leads us to consider the action with the strong constraints. Otherwise, we do not have gauge invariant action. When we use the strong constraints, the non-gauge invariant terms can be annihilated. Due to the manifest $O(D, D)$ structure,  we rewrite the weak constraint as
\bea
\partial^A\partial_A(\mbox{field})=0,
\eea
where $\partial_A$ is defined by
\be
{\partial_A }\equiv
 \begin{pmatrix} \,\tilde{\partial}^M \, \\[0.6ex] {\partial_M } \end{pmatrix}
\ee
and $\partial^A\equiv\eta^{AB}\partial_C$. The index $A=0, 1\cdots, 2D-1$ (We denote the double target indices from $A$ to $K$.). We use $\eta$ to raise and lower the indices for the $O(D, D)$ tensors
\be
 h= \begin{pmatrix} a& b \\ c& d \end{pmatrix}
\, ,
\qquad   h^t \eta h = \eta, \qquad\eta= \begin{pmatrix} 0& I \\ I& 0 \end{pmatrix},
\ee
 where $a$, $b$, $c$ and $d$ are $D$ by $D$ matrices. We define the transpose of $h$ as $h^t$. 
We use $X^A$ to combine the ordinary and dual coordinates by
\be
{X^A}\equiv
 \begin{pmatrix} \,\tilde x_M\, \\[0.6ex] {x^M} \end{pmatrix}.
\ee

\subsection{Gauge Transformation}
We review the gauge transformation and introduce the generalized Lie derivative, $C$-bracket, and $D$-bracket for the generalized metric formulation \cite{Hohm:2010pp}. In the end of this section, we show that the Courant and Dorfman brackets can be obtained from the $C$ and $D$-brackets by using the strong constraints.

The gauge transformation is
\bea
\label{gauge tran_BI}
\delta{\cal E}_{MN}&\equiv&\delta(g+B)_{MN}
\nn\\
&=&{\cal D}_M\tilde{\xi}_N-\bar{{\cal D}}_N\tilde{\xi}_M
\nn\\
&&+\xi^P\partial_P{\cal E}_{MN}+{\cal D}_M\xi^P{\cal E}_{PN}+\bar{{\cal D}}_{N}\xi^P{\cal E}_{MP},
\nn\\
\delta d&=&-\frac{1}{2}\partial_P\xi^P+\xi^P\partial_P d,
\eea
where
\bea
e^{-2d}=\sqrt{-\det{g}}e^{-2\phi},
\eea
\bea
{\cal D}_M&=&\partial_M-{\cal E}_{MN}\tilde{\partial}^N, 
\nn\\
\bar{{\cal D}}_M&=&\partial_M+{\cal E}_{NM}\tilde{\partial}^N,
\eea
 and $\phi$ is the dilaton. Then we introduce the generalized metric (${\cal H}_{AB}$),
   \be
  {\cal H}~ \equiv ~ {\cal H}^{\bullet\,\bullet}  \,,
  \ee
\be
  {\cal H} \ = \
  \begin{pmatrix}    g-Bg^{-1}B & Bg^{-1}\\[0.5ex]
  -g^{-1}B & g^{-1}\end{pmatrix}\, .
 \ee
This matrix is a symmetric matrix with the $O(D, D)$ symmetry,
 \be
 {\cal H}\,\eta \,{\cal H}=\eta\;.
  \ee
The inverse of ${\cal H}$ is
\be
{\cal H}^{-1} =  \eta {\cal H} \eta\,,
\ee
\bea
  {\cal H}^{-1}~ &\equiv& ~ {\cal H}_{\bullet\,\bullet}  \,\ = \ \left({\cal H}^{AB}\right)^{-1} \
\nn\\
 &=& \
  \begin{pmatrix}    g^{-1} & -g^{-1}B\\[0.5ex]
  Bg^{-1} & g-Bg^{-1}B\end{pmatrix}\;.
  \eea
  The gauge transformation of the generalized metric is
\bea
\delta_{\xi} {\cal H}^{AB}&=&\xi^P\partial_P{\cal H}^{AB}+(\partial^A\xi_C-\partial_C\xi^A){\cal H}^{CB}
\nn\\
&&+(\partial^B\xi_C-\partial_P\xi^B){\cal H}^{AC},
\eea
where
\be
{\xi^A}\equiv
 \begin{pmatrix} \,\tilde{\xi}_M\, \\[0.6ex] {\xi^M} \end{pmatrix}.
\ee
Then we define the generalized Lie derivative from 
\bea
\hat{\cal L}_{\xi}{\cal H}^{AB}\equiv\delta_{\xi}{\cal H}^{AB},
\eea
which satisfies the Leibniz rule. The generalized Lie derivative acting on the constant metric ($\eta$) \emph{is} zero, but the ordinary Lie derivative is not.
The gauge algebra is closed by imposing the strong constraints as
\bea
[\hat{\cal L}_{\xi_1}, \hat{\cal L}_{\xi_2}]=\hat{\cal L}_{[\xi_1, \xi_2]_C},
\eea
where the $C$-bracket is defined by
\bea
[\xi_1, \xi_2]_C^A&=&\xi_1^C\partial_C\xi_2^A-\xi_2^C\partial_C\xi_1^A-\frac{1}{2}\eta^{AC}\eta_{DE}\xi_1^D\partial_C\xi_2^E
\nn\\
&&+\frac{1}{2}\eta^{AC}\eta_{DE}\xi_2^D\partial_C\xi_1^E.
\eea
We define the $D$-bracket in the case of the generalized vector as
\bea
[A, B]_D\equiv\hat{\cal L}_A B.
\eea
The difference between the $C$-bracket and $D$-bracket is a total derivative term
\bea
[A, B]_D^A=[A, B]_C^A+\frac{1}{2}\partial^A(B^C A_C).
\eea
Now, we assume that all parameters are independent of $\tilde{x}$ to get the Courant bracket from the $C$-bracket. Then we obtain
\bea
\lbrack\xi_1, \xi_2\rbrack_C^M&=&\xi_1^P\partial_P\xi_2^M-\xi_2^P\partial_P\xi_1^M
\nn\\
&&=({\cal L}_{\xi_1}\xi_2)^M\equiv([\xi_1, \xi_2])^M,
\nn\\
\lbrack\xi_1, \xi_2\rbrack_{CM}&=&\xi^P_1\partial_P\tilde{\xi}_{2M}-\xi^P_2\partial_P\tilde{\xi}_{1M}
\nn\\
&&-\frac{1}{2}(\xi_1^P\partial_M\tilde{\xi}_{2P}-\tilde{\xi}_{2P}\partial_M\xi^P_1)
\nn\\
&&+\frac{1}{2}(\xi_2^P\partial_M\tilde{\xi}_{1P}-\tilde{\xi}_{1P}\partial_M\xi^P_2)
\nn\\
&=&
\xi^P_1\partial_P\tilde{\xi}_{2M}-\xi^P_2\partial_P\tilde{\xi}_{1M}
+(\partial_M\xi_1^P)\tilde{\xi}_{2P}
\nn\\
&&-\frac{1}{2}\partial_M(\xi_1^P\tilde{\xi}_{2P})
-(\partial_M\xi_2^P)\tilde{\xi}_{1P}
\nn\\
&&+\frac{1}{2}\partial_M(\xi_2^P\tilde{\xi}_{1P})
\nn\\
&=&\bigg({\cal L}_{\xi_1}\tilde{\xi}_2-\frac{1}{2}d(i_{\xi_1}\tilde{\xi}_2)\bigg)_M
\nn\\
&&-\bigg({\cal L}_{\xi_2}\tilde{\xi}_1-\frac{1}{2}d(i_{\xi_2}\tilde{\xi}_1)\bigg)_M.
\eea
This is exactly the same as 
\bea
[A+\alpha, B+\beta]_{\mbox{Cour}}&=&[A, B]+{\cal L}_A\beta-{\cal L}_B\alpha
\nn\\
&&-\frac{1}{2}d(i_A\beta-i_B\alpha),
\eea
where $A$, $B$ are vectors, and $\alpha$, $\beta$ are one-forms.
Similarly, we also obtain the Dorfman bracket \cite{Gualtieri:2003dx, Hitchin:2004ut} 
\bea
[A+\alpha, B+\beta]_{\mbox{Dor}}=[A, B]+{\cal L}_A\beta-i_Bd\alpha
\nn\\
\eea
from the $D$-bracket. We express the Dorfman bracket in a different way instead of the conventional $(A+\alpha)\circ(B+\beta)$ with the consistent notation. The $D$-bracket has the Jacobi identity
\bea
[A, [B, C]_D]_D=[[A, B]_D, C]_D+ [B, [A, C]_D]_D,
\nn\\
\eea
 but it is not antisymmetric. For the $C$-bracket, it does not satisfy the Jacobi identity, but it is antisymmetric. In other words, the $C$ and $D$-brackets are \emph{not} the Lie brackets.

\section{Review of the D-brane Theory}
\label{3}
The well-known D-brane theory comes from a two dimensional worldsheet theory with the Dirichlet and the Neumann boundary conditions. We start from the bulk action
\bea
&&\frac{1}{2}\int d^2\sigma\sqrt{\det{(-h_{\gamma\delta})}}\ \bigg(h^{\alpha\beta}\partial_{\alpha}X^Mg_{MN}\partial_{\beta}X^N
\nn\\
&&-\epsilon^{\alpha\beta}\partial_{\alpha}X^MB_{MN}\partial_{\beta}X^N+R^{(2)}\phi\bigg),
\eea
where $R^{(2)}$ is the worldsheet two dimensional Ricci scalar, $h_{\alpha\beta}$ is the worldsheet metric, $\epsilon^{01}=1$, and $\alpha=0, 1$ (We use the Greek letters to indicate the worldsheet indices.). Based on the diffemorphism and Weyl symmetry, we choose $h_{\alpha\beta}=(-, +)$. The gauge symmetries on the target space are the diffemorphism and one-form gauge transformation. The one-form gauge transformation ($\delta_{\mbox{one-form}} B_{MN}=\partial_M\Lambda_N-\partial_N\Lambda_M$) gives 
\bea
-\int d^2 \sigma\ \epsilon^{\alpha\beta}\partial_{\alpha}\bigg(\Lambda_N\partial_{\beta}X^N\bigg).
\eea
If we consider that all fields vanish at infinity, this term should vanish. Then we have the gauge invariance. In this case, we do not have the open string theory. When we choose the Neumann boundary condition in the $\sigma^1$ direction, this term will be the non-gauge invariant term. Nevertheless, we add a boundary term to cancel this non-gauge invariant term to let this theory to be gauge invariant. At first, we integrate out the $\sigma^1$ direction, then we can obtain
\bea
\int d\sigma^0\ \Lambda_N\partial_0X^N
\eea
on the boundary. We should add the boundary term to cancel it. The boundary term is
\bea
-\int d\sigma^0 A_N\partial_0X^N.
\eea
The one-form gauge transformation of $A_M$ is 
\bea
\delta_{\mbox{one-form}}A_{M}=\Lambda_M
\eea
so the gauge transformation of the boundary term is
\bea
-\int d\sigma^0\ \Lambda_N\partial_0X^N.
\eea
We can obtain gauge invariant action when we consider the Neumann boundary condition in the $\sigma^1$ direction. The one-loop $\beta$ calculation \cite{Callan:1986bc} of this two dimensional string sigma model shows the non-trivial DBI term
\bea
\int dx\ e^{-\phi} \sqrt{-\det{(g+B-F)}},
\eea
where
\bea
F_{MN}=\partial_MA_N-\partial_NA_M.
\eea
Now, we review the T-duality of the DBI term. The T-duality rules of the open string are given by
\bea
g^{\prime}_{yy}=\frac{1}{g_{yy}}, \qquad g^{\prime}_{ya}=\frac{B_{ya}}{g_{yy}},
\nn\\
\eea
\bea
 g^{\prime}_{ab}=g_{ab}-\frac{g_{ya}g_{yb}-B_{ya}B_{yb}}{g_{yy}},
\nn\\
\eea
\bea
B^{\prime}_{ya}=\frac{g_{ya}}{g_{yy}}, \qquad B^{\prime}_{ab}=B_{ab}-\frac{g_{ya}B_{yb}-B_{ya}g_{yb}}{g_{yy}}, 
\nn\\
\eea
\bea
\phi^{\prime}=\phi-\ln\sqrt{g_{yy}},
\nn\\
\eea
where $g$, $B$ and $\phi$ are the original fields, and $g^{\prime}$, $B^{\prime}$ and $\phi^{\prime}$ are the fields after performing the T-dual transformation. We specify the spacetime directions (non-compact directions) by $a$  and the compact directions by $y$ (We denote the non-compact target indices from $a$ to $h$ and the compact target indices from $i$ to $z$.). We perform the dimensional reduction when we consider the T-duality of the open string theory. The reduction rule is given by
\bea
F^{\prime}_{ab}=F_{ab}, \qquad F^{\prime}_{ay}=\partial_a\Phi, \qquad F^{\prime}_{yz}=0,
\nn\\
\eea
where $\Phi$ is the scalar field that lives on the compact directions and $F^{\prime}$ is the dual field strength. A useful identity during the derivation of the T-dual in the DBI theory is
\bea
\det\bigg(W_{MN}\bigg)=W_{yy}\det\bigg(W_{ab}-\frac{W_{ay}W_{yb}}{W_{yy}}\bigg).
\nn\\
\eea
When we identify $W=g+B-F$, we will obtain
\bea
&&\det\bigg((g+B-F)_{MN}\bigg)=g_{yy}\det\bigg((g+B-F)_{ab}
\nn\\
&&-\frac{(g+B-F)_{ay}(g+B-F)_{yb}}{g_{yy}}\bigg).
\eea
Rewriting the original fields in terms of the dual fields by the T-dual operations, then we obtain
\bea
&&\bigg(g^{\prime}+B^{\prime}-F\bigg)_{ab}=(g+B-F)_{ab}
\nn\\
&&-\frac{g_{ya}g_{yb}-B_{ya}B_{yb}-g_{ya}B_{yb}+B_{ya}g_{yb}}{g_{yy}}
\nn\\
&=&(g+B-F)_{ab}-\frac{g_{ya}(g_{yb}-B_{yb})+B_{ya}(g_{yb}-B_{yb})}{g_{yy}}
\nn\\
&=&(g+B-F)_{ab}-\frac{(g+B)_{ay}(g+B)_{yb}}{g_{yy}},
\nn\\
\eea
\bea
&&\frac{1}{g_{yy}}\partial_a\Phi\partial_b\Phi+\partial_a\Phi(g^{\prime}+B^{\prime})_{yb}+\partial_b\Phi(g^{\prime}-B^{\prime})_{ya}
\nn\\
&=&\frac{1}{g_{yy}}\partial_a\Phi\partial_b\Phi+\partial_a\Phi\bigg(\frac{g_{yb}+B_{yb}}{g_{yy}}\bigg)
\nn\\
&&+\partial_b\Phi\bigg(\frac{g_{ya}-B_{ya}}{g_{yy}}\bigg),
\nn\\
\eea
\bea
e^{-\phi}\sqrt{g_{yy}}=e^{-(\phi-\ln\sqrt{g_{yy}})}=e^{-\phi^{\prime}}
\eea
and
\bea
P(g^{\prime}+B^{\prime})_{ab}&\equiv&(g^{\prime}+B^{\prime})_{ab}+\partial_a\Phi(g^{\prime}
+B^{\prime})_{yb}
\nn\\
&&+\partial_b\Phi(g^{\prime}+B^{\prime})_{ay}
+g^{\prime}_{yy}\partial_a\Phi\partial_b\Phi.
\nn\\
\eea
Therefore, we get
\bea
&&e^{-\phi}\sqrt{-\det(g+B-F)_{MN}}
\nn\\
&=&e^{-\phi^{\prime}}\sqrt{-\det\bigg({P(g^{\prime}+B^{\prime})_{ab}-F_{ab}\bigg)}}.
\eea
The DBI term has the T-dual invariant form. However, the T-duality of the open string is not exactly the same as the T-duality of the closed string because the T-duality of the open string  changes dimensions of spacetime. For example, the scalar dilaton ($d$) is an invariant quantity in the massless closed string theory, but the scalar dilaton in the open string is not. Even if we lose the T-dual invariant quantities, we still have the invariant form based on the T-duality in the open string theory.
\section{Double Field Theory of the DBI Model}
\label{4}
We introduce our set-ups and notations for the double field theory of the DBI model. Then we write down the gauge transformation without the double coordinates. We also show gauge invariance. At the end of this section, we construct the gauge transformation in the double field theory and define the $F$-bracket. We perform the B-transformation on the $F$-bracket with the strong constraints to compare with the Courant bracket.

\subsection{Set-Ups and Notations}
We define our notations for the double field theory of the DBI model and construct the double field theory of the DBI model by combining the ordinary coordinates with the dual coordinates. The ordinary coordinates are associated with the Neumann boundary conditions and the dual coordinates are associated with the Dirichlet boundary conditions. Our notations are given by
\bea
{\Lambda_M }\equiv
 \begin{pmatrix} \,\epsilon^i \, \\[0.6ex] {\Lambda_a } \end{pmatrix}, \qquad
{\epsilon^M }\equiv
 \begin{pmatrix} \,\Lambda_i \, \\[0.6ex] {\epsilon^a } \end{pmatrix}, \qquad
{A_M }\equiv
 \begin{pmatrix} \,\phi^i \, \\[0.6ex] {A_a } \end{pmatrix}, 
\nn\\
\eea
\bea
{\partial_M }&\equiv&
 \begin{pmatrix} \, \tilde{\partial^i} \, \\[0.6ex] {\partial_a } \end{pmatrix}, \qquad
 {\tilde{\partial}^M }\equiv
 \begin{pmatrix} \, \partial_i \, \\[0.6ex] {\tilde{\partial}^a } \end{pmatrix},
 \nn\\
 \eea
 where the indices $a=0, 1, \cdots, p$ and $i=(p+1), (p+2), \cdots, (D-1)$ in the D$p$-brane theory. The index $a$ denotes the parallel (world-volume) directions and index $i$ denotes the transversal
 directions. If we perform the dimensional reduction on $i$ directions, this is equivalent to using
\bea
\tilde{\partial}^i(\mbox{field})=0.
\eea
The T-duality rules of the background fields can be manifestly obtained from
\bea
{\cal E}^{\prime}(X^{\prime})=(a{\cal E}(X)+b)(c{\cal E}(X)+d)^{-1},
\nn\\
\eea
\bea
 d^{\prime}(X^{\prime})=d(X),\qquad X^{\prime}=hX,\qquad {\cal E}\equiv g+B,
\nn\\
\eea
where
\be
 h= \begin{pmatrix} a& b \\ c& d \end{pmatrix}
\, ,
\qquad   h^t \eta h = \eta, \qquad\eta= \begin{pmatrix} 0& I \\ I& 0 \end{pmatrix}.
\ee
The T-duality of the open string should change dimensions of spacetime, but the manifest T-duality rules of the closed string theory does not change dimensions. In other words, we lose the meaning of the manifest T-duality at the level of action, but the meaning of the manifest T-duality still remains at the level of transformation. The main reason is due to the fact that the dimensional reduction changes dimensions. We will use the gauge transformation to explain that the $O(D, D)$ structure is not suitable to describe the open string theory. 

We use $\tilde{\partial}^M(\mbox{field})=0$ and $\tilde{\partial}^i(\mbox{field})=0$ to guarantee gauge invariance. From ${\cal E}^{\prime}(X^{\prime})=(a{\cal E}(X)+b)(c{\cal E}(X)+d)^{-1}$ with a particular choice of the $O(D, D)$ element ($h$), we can get the Buscher's rule.
The convention for $X$ is
\bea
X^M= \begin{pmatrix} \,\tilde{X}_i \, \\[0.6ex] {X^a } \end{pmatrix}, \qquad 
\tilde{X}_M= \begin{pmatrix} \,X_i \, \\[0.6ex] {\tilde{X}^a } \end{pmatrix},
\nn\\
\eea
\bea
X^A= \begin{pmatrix} \,\tilde{X}_M \, \\[0.6ex] {X^M } \end{pmatrix}\equiv X.
\eea
We use $\eta$ to define $X_A\equiv\eta_{AB}X^B$.
From the above discussion, we can show that the exchange of the coordinates is equivalent to performing the T-duality rules. We define a new element related to ${\cal E}$. This new element ($t_{MN}$) is based on the T-dual operation.
\bea
t_{ab}\equiv{\cal E}_{ab}-{\cal E}_{ak}{\cal E}^{kl}{\cal E}_{lb},
\qquad
t^i{}_b\equiv{\cal E}^{ik}{\cal E}_{kb},
\nn
\eea
\bea
t_a{}^j\equiv-{\cal E}_{ak}{\cal E}^{kj},
\qquad
t^{ij}\equiv{\cal E}^{ij}.
\nn
\eea
We use 
\bea
 t_{MN}= \begin{pmatrix} t^{ij}& t^i{}_b \\ t_a{}^j& t_{ab} \end{pmatrix}
\eea
to combine all new elements.
If we consider the D($D$-1)-brane theory, we have $t_{MN}={\cal E}_{MN}=(g+B)_{MN}$. For convenience, we define $t_{MN}\equiv s_{MN}+a_{MN}$, where $s\equiv\frac{t+t^t}{2}$ and $a\equiv\frac{t-t^t}{2}$. We embed the Buscher's rule in the
\bea
t^{\prime}(X^{\prime})=(at(X)+b)(ct(X)+d)^{-1}
\eea
with a particular $O(D, D)$ element ($h$) by choosing $a$, $b$, $c$ and $d$.

\subsection{Gauge Transformation}
We first write the gauge transformation of the DBI theory:
\bea
\delta t_{MN}
&\equiv&\partial_M\Lambda_N-\partial_N\Lambda_M+{\cal L}_{\epsilon}t_{MN}+{\cal L}_{\Lambda}t_{MN}
\nn\\
&&+t_{MQ}(\tilde{\partial}^P\epsilon^Q-\tilde{\partial}^Q\epsilon^P)t_{PN},
\nn\\
\delta s_{MN}&=&{\cal L}_{\epsilon}s_{MN}+{\cal L}_{\Lambda}s_{MN}
\nn\\
&&+(s_{MQ}a_{NP}+s_{NQ}a_{MP})(\tilde{\partial}^Q\epsilon^P-\tilde{\partial}^P\epsilon^Q),
\nn\\
\delta a_{MN}&=&\partial_M\Lambda_N-\partial_N\Lambda_M+{\cal L}_{\epsilon}a_{MN}
+{\cal L}_{\Lambda}a_{MN}
\nn\\
&&-s_{MP}(\tilde{\partial}^P\epsilon^Q-\tilde{\partial}^Q\epsilon^P)s_{QN}
\nn\\
&&-a_{MP}(\tilde{\partial}^P\epsilon^Q-\tilde{\partial}^Q\epsilon^P)a_{QN},
\nn\\
\eea
where
\bea
{\cal L}_{\epsilon}t_{MN}&=&\epsilon^Q\partial_Qt_{MN}+(\partial_M\epsilon^Q)t_{QN}
+t_{MQ}\partial_N\epsilon^Q,
\nn\\
{\cal L}_{\Lambda}t_{MN}&=&\Lambda_Q\tilde{\partial}^Qt_{MN}
+(\tilde{\partial}^Q\Lambda_M)t_{NQ}-\tilde{\partial}^Q\Lambda_Nt_{MQ}.
\nn\\
\eea
The gauge transformation of field strength is
\bea
\delta F_{MN}=\partial_M\Lambda_N-\partial_N\Lambda_M+{\cal L}_{\epsilon}F_{MN}.
\eea
From the gauge transformation of the field strength, we show that the DBI theory is gauge invariant with $\tilde{\partial}^M$=0.
We use some useful matrix identities to rewrite the DBI action to show gauge invariance.  Now we decompose $t$ as
\bea
 t&=& \begin{pmatrix} \delta_{ac}& -{
\cal E}_{ak}{\cal E}^{kl} \\ 0& {\cal E}^{il} \end{pmatrix}
\begin{pmatrix} {\cal E}_{cb}& 0 \\ {\cal E}_{lb}& \delta_{lj} \end{pmatrix}
\nn\\
&=& \begin{pmatrix} \delta_{ac}& {\cal E}_{ak} \\ 0& {\cal E}_{ik} \end{pmatrix}^{-1}
\begin{pmatrix} {\cal E}_{cb}& 0 \\ {\cal E}_{kb}& \delta_{kj} \end{pmatrix}.
\eea
We define $m\equiv \begin{pmatrix} \delta_{ac}& {\cal E}_{ak} \\ 0& {\cal E}_{ik} \end{pmatrix}$ and $n\equiv \begin{pmatrix} {\cal E}_{cb}& 0 \\ {\cal E}_{kb}& \delta_{kj} \end{pmatrix}$.
Therefore, we obtain
\bea
s&=&\frac{1}{2}\bigg(m^{-1}n+n^t(m^{-1})^t\bigg)
\nn\\
&&=\frac{1}{2}m^{-1}(nm^t+mn^t)(m^{-1})^t, 
\nn\\
\eea
\bea
mn^t= \begin{pmatrix} {\cal E}_{ac}& 2g_{ak}\\ 0& {\cal E}_{ik} \end{pmatrix}, \qquad nm^t= \begin{pmatrix} {\cal E}_{ca}& 0\\ 2g_{ka}& {\cal E}_{ki} \end{pmatrix}.
\nn\\
\eea
Hence, we show
\bea
s=m^{-1}g(m^{-1})^t.
\eea
This immediately implies
\bea
\det s=(\det m)^{-2}\det g=(\det t^{ij})^2\det g.
\nn\\
\eea
Then we decompose $\det t$ as
\bea
\det t&=&\bigg(\det (s+a)\det(s+a)\bigg)^{\frac{1}{2}}
\nn\\
&=&\bigg(\det s\det(1+s^{-1}a)\det(1+as^{-1})\det s\bigg)^{\frac{1}{2}}
\nn\\
&=&\det s\bigg(\det(1+s^{-1}a)\det(1-s^{-1}a)\bigg)^{\frac{1}{2}}
\nn\\
&=&\det s\bigg(\det(1-s^{-1}as^{-1}a)\bigg)^{\frac{1}{2}}
\nn\\
&=&(\det s)^{\frac{1}{2}}\bigg(\det( s-as^{-1}a)\bigg)^{\frac{1}{2}}.
\eea
In the case of $t_F=t-F$, we get a similar result as
\bea
&&\det t_F
\nn\\
&=&(\det s)^{\frac{1}{2}}\bigg[\det\bigg(s-(a-F)s^{-1}(a-F)\bigg)\bigg]^{\frac{1}{2}} .
\nn\\
\eea
By using
\bea
\det\begin{pmatrix} A& B\\ C& D \end{pmatrix}=\det\bigg(A-BD^{-1}C\bigg)\det D,
\nn\\
\eea
we obtain
\bea
\det t_F&=&\det {\cal E}^{ij}\det \bigg(P({\cal E})-F\bigg)
\nn\\
&=&\det t^{ij}\det \bigg(P({\cal E})-F\bigg).
\eea
Then we get
\bea
&&\bigg[-\det\bigg(P({\cal E})-F\bigg)\bigg]^{\frac{1}{2}}=(-\det t_F)^{\frac{1}{2}}\frac{1}{(\det t^{ij})^{\frac{1}{2}}}
\nn\\
&=&(-\det s)^{\frac{1}{4}}\frac{1}{(\det t^{ij})^{\frac{1}{2}}}
\nn\\
&&\times\bigg[\det\bigg(s-(a-F)s^{-1}(a-F)\bigg)\bigg]^{\frac{1}{4}}
\nn\\
&=&(-\det g)^{\frac{1}{4}}\bigg[\det\bigg(s-(a-F)s^{-1}(a-F)\bigg)\bigg]^{\frac{1}{4}}
\nn\\
\eea
 and the gauge transformation of $\det g$
\bea
\delta(\det g)&=&(\det g) g^{-1}\delta g
\nn\\
&=&(\det g) g^{ab}\bigg(\epsilon^c\partial_c g_{ab}+\partial_a\epsilon^c g_{cb}+\partial_b\epsilon^cg_{ca}\bigg)
\nn\\
&=&\epsilon^c\partial_c\det g+2\partial_c\epsilon^c\det g.
\eea
Similarly, we also get
\bea
&&\delta\bigg[\det\bigg(s-(a-F)s^{-1}(a-F)\bigg)\bigg]
\nn\\
&=&\epsilon^c\partial_c\bigg[\det\bigg(s-(a-F)s^{-1}(a-F)\bigg)\bigg]
\nn\\
&&+2\partial_c\epsilon^c\bigg[\det\bigg(s-(a-F)s^{-1}(a-F)\bigg)\bigg].
\nn\\
\eea
Hence, we have
\bea
&&\delta\bigg[\bigg(-\det g\bigg)^{\frac{1}{4}}\bigg]
\nn\\
&=&\epsilon^c\partial_c\bigg(-\det g\bigg)^{\frac{1}{4}}+\frac{1}{2}\partial_c\epsilon^c\bigg(-\det g\bigg)^{\frac{1}{4}}
\nn\\
\eea
and
\bea
&&\delta\bigg[\det\bigg(s-(a-F)s^{-1}(a-F)\bigg)^{\frac{1}{4}}\bigg]
\nn\\
&=&\epsilon^c\partial_c\det\bigg(s-(a-F)s^{-1}(a-F)\bigg)^{\frac{1}{4}}
\nn\\
&&+\frac{1}{2}\partial_c\epsilon^c\det\bigg(s-(a-F)s^{-1}(a-F)\bigg)^{\frac{1}{4}}.
\nn\\
\eea
Then we use the above gauge transformation to obtain
\bea
&&\delta\bigg[-\det\bigg(P({\cal E})-F\bigg)\bigg]^{\frac{1}{2}}
\nn\\
&=&\partial_c\bigg\{\epsilon^c\bigg[-\det\bigg(P({\cal E})-F\bigg)\bigg]^{\frac{1}{2}}\bigg\}.
\eea
Since the gauge transformation of the dilation is $\delta\phi=\epsilon^c\partial_c\phi$, we have gauge invariance for the DBI action. This is easy to deduce that the gauge transformation of the scalar dilaton is
\bea
\delta d=\epsilon^M\partial_M d-\frac{1}{2}\partial_M\epsilon^M.
\eea
We rewrite this theory by using $d$, $t$ and $F$ as well.
The Lagrangian becomes
\bea
e^{-d}\bigg(-\det( t-F )\bigg)^{\frac{1}{2}}\frac{1}{(-\det \frac{t+t^t}{2})^{\frac{1}{4}}}.
\eea
However, this Lagrangian does not have the $O(D, D)$ structure.  Later we will discuss more about this issue. This setup is based on the generalized geometry \cite{Asakawa:2012px}. We provide a way to extend from the generalized geometry to the double formulation.
\subsection{Bracket}
We discuss what kind of the bracket that appears in the double field theory of the DBI theory. Since the gauge transformation of the DBI theory without the one-form gauge field is the same as the gauge transformation of the massless closed string theory, we have the Courant bracket in this theory. If we include the gauge field, we will obtain the $F$-bracket. We start from the gauge transformation of the gauge field
\bea
\delta_2 A_M&=&\Lambda_{2M}+\epsilon_2^NF_{NM},
\eea
\bea
\delta_1\delta_2A_M&=&\epsilon_2^N\delta_1F_{NM}
\nn\\
&=&\epsilon_2^N\bigg(\partial_N
\Lambda_{1M}
-\partial_M\Lambda_{1N}+\epsilon_1^P\partial_PF_{NM}
\nn\\
&&+(\partial_N\epsilon_1^P)F_{PM}
+(\partial_M\epsilon_1^P)F_{NP}\bigg),
\eea
\bea
\lbrack\delta_1, \delta_2\rbrack A_M&=&\epsilon_2^N\bigg(\partial_N
\Lambda_{1M}
-\partial_M\Lambda_{1N}\bigg)
\nn\\
&&-\epsilon_1^N\bigg(\partial_N
\Lambda_{2M}
-\partial_M\Lambda_{2N}\bigg)
\nn\\
&&+\epsilon_2^N\epsilon_1^P\partial_PF_{NM}+\epsilon_2^N(\partial_N\epsilon_1^P)F_{PM}
\nn\\
&&
+\epsilon_2^N(\partial_M\epsilon_1^P)F_{NP}
\nn\\
&&-\epsilon_1^N\epsilon_2^P\partial_PF_{NM}-\epsilon_1^N(\partial_N\epsilon_2^P)F_{PM}
\nn\\
&&
-\epsilon_1^N(\partial_M\epsilon_2^P)F_{NP},
\eea
\bea
&&\epsilon_2^N\epsilon_1^P\partial_PF_{NM}
-\epsilon_1^N\epsilon_2^P\partial_PF_{NM}
\nn\\
&=&\epsilon_2^N\epsilon_1^P\bigg(\partial_PF_{NM}-\partial_NF_{PM}\bigg)
\nn\\
&=&\epsilon_2^N\epsilon_1^P\bigg(-\partial_P\partial_MA_N+\partial_N\partial_MA_P\bigg)
\nn\\
&=&\epsilon_2^N\epsilon_1^P\partial_MF_{NP}.
\eea
Therefore, we have
\bea
\epsilon^{\prime M}&=&\epsilon_1^N\partial_N\epsilon_2^M-\epsilon_2^N\partial_N\epsilon_1^M,
\nn\\
\Lambda^{\prime}_M&=&\epsilon_1^N\bigg(\partial_N\Lambda_{2M}
-\partial_M\Lambda_{2N}
\bigg)
\nn\\
&&-\epsilon_2^N\bigg(\partial_N\Lambda_{1M}-\partial_M\Lambda_{1N}\bigg)
\nn\\
&&-\partial_M\bigg(\epsilon_2^N\epsilon_1^PF_{NP}
\bigg),
\nn\\
\lbrack\delta_1,\delta_2\rbrack A_M&=&-\delta^{\prime}A_M.
\eea
When we use the double indices to rewrite these parameters, we need to double the gauge field to do contraction. We would not like to see this situation because this makes the case of the double gauge fields unavoidable. Doubling the gauge field makes a theory more difficult to be described and this is not the double field theory that we consider. In order to remove this field dependence, we need to redefine the gauge transformation of the gauge field without changing the gauge transformation of the field strength. The new gauge transformation is given by
\bea
\delta A_M&=&\Lambda_M+\partial_M(\epsilon^NA_N)+\epsilon^NF_{NM}
\nn\\
&=&\Lambda_M+{\cal L}_{\epsilon}A_M.
\eea 
Then we get
\bea
\label{gp}
\epsilon^{\prime M}&=&\epsilon_1^N\partial_N\epsilon_2^M-\epsilon_2^N\partial_N\epsilon_1^M,
\nn\\
\Lambda^{\prime}_M&=&\epsilon_1^{N}\partial_{N}\Lambda_{2M}
+(\partial_M\epsilon_1^N)\Lambda_{2N}-\epsilon_2^{N}\partial_{N}\Lambda_{1M}
\nn\\
&&-(\partial_M\epsilon_2^N)\Lambda_{1N}
\nn\\
&=&{\cal L}_{\epsilon_1}\Lambda_{2M}-{\cal L}_{\epsilon_2}\Lambda_{1M}.
\eea
We also define a new bracket from
\bea
\lbrack\xi_1,\xi_2\rbrack_F^A&=&\bigg(\xi_1^D\partial_D\xi_2^A-\xi_2^D\partial_D\xi_1^A\bigg)
\nn\\
&&-\frac{1}{2}\bigg(\xi_1^D\partial^A\xi_{2D}-\xi_2^D\partial^A\xi_{1D}\bigg)
\nn\\
&&-\frac{1}{2}\partial^A\bigg(\xi_{2D}Z^{D}{}_E\xi_1^E\bigg),
\nn\\
\eea
where
\bea
Z\equiv Z^{A}{}_B\equiv\begin{pmatrix} -1 & 0
 \\ 0 & 1  \end{pmatrix}.
\eea
If we use the strong constraints, we get (\ref{gp}) consistently. In other words, we obtain
\bea
\lbrack\xi_1,\xi_2\rbrack_F^M=\epsilon^{\prime M}, \qquad \lbrack\xi_1,\xi_2\rbrack_{FM}=\Lambda^{\prime}_M
\eea
with the strong constraints.
We note that $Z$ is not an $O(D,D)$ matrix. But we still use $\eta$ to raise or lower indices.
This is easy to deduce
\bea
\lbrack\delta_1, \delta_2\rbrack=-\delta_{\lbrack\xi_1, \xi_2\rbrack_F}.
\eea
If we do not use the language of the double field theory to describe the D-brane theory, the gauge transformation of the gauge field has the ambiguity. When considering the double field theory, this ambiguity will be removed. This implies that the double field theory has more constraints to restrict a theory to construct the action and find the gauge transformation. 

We would like to know whether the property of the automorphism exists after we perform the B-transformation. The B-transformation is
\bea
e^B\equiv\begin{pmatrix} 1& 0\\ B& 1 \end{pmatrix}, 
\eea
\bea
 e^B\begin{pmatrix} X\\ \xi \end{pmatrix}=\begin{pmatrix} X\\ \xi+BX \end{pmatrix}=\begin{pmatrix} X\\ \xi+i_XB \end{pmatrix}.
\eea
We first calculate the Courant bracket
\bea
&&\lbrack e^B(X+\xi), e^B(Y+\eta)\rbrack_{\mbox{Cour}}
\nn\\
&=&\lbrack X+\xi+i_X B, Y+\eta+i_Y B\rbrack_{\mbox{Cour}}
\nn\\
&=&\lbrack X+\xi, Y+\eta\rbrack_{\mbox{Cour}}+\lbrack X, i_Y B\rbrack_{\mbox{Cour}}
\nn\\
&&+\lbrack i_X B, Y\rbrack_{\mbox{Cour}}
\nn\\
&=&\lbrack X+\xi, Y+\eta\rbrack_{\mbox{Cour}}+{\cal L}_X i_Y B
\nn\\
&&-\frac{1}{2}di_Xi_Y B-{\cal L}_Yi_X B+\frac{1}{2}di_Yi_X B
\nn\\
&=&\lbrack X+\xi, Y+\eta\rbrack_{\mbox{Cour}} +i_{\lbrack X, Y\rbrack}B+i_Yi_XdB
\nn\\
&=&e^B\bigg(\lbrack X+\xi, Y+\eta\rbrack_{\mbox{Cour}}\bigg)+i_Yi_XdB.
\eea
If $dB=0$, we get the automorphism after using the B-transformation. This shows that this theory can define a $H$-flux ($dH=0$) and possibly be extended to be described by the $O(D, D)$ structure. For the massless closed string theory with the double formulation, we use the $O(D, D)$ structure to represent this theory with the $H$-flux. We include the gauge field to discuss the $F$-bracket. For a convenience, we define a notation for the $F$-bracket with the strong constraints
\bea
\lbrack X+\xi, Y+\eta\rbrack_{F}=\lbrack X, Y\rbrack+{\cal L}_{X}\eta-{\cal L}_{Y}\xi.
\nn\\
\eea
Then examination of the automorphism is given by
\bea
&&\lbrack e^B(X+\xi), e^B(Y+\eta)\rbrack_F
\nn\\
&=&\lbrack X+\xi+i_X B, Y+\eta+i_Y B\rbrack_F
\nn\\
&=&\lbrack X+\xi, Y+\eta\rbrack_F+\lbrack X, i_Y B\rbrack_F+\lbrack i_X B, Y\rbrack_F
\nn\\
&=&\lbrack X+\xi, Y+\eta\rbrack_F+{\cal L}_X i_Y B-{\cal L}_Yi_X B
\nn\\
&=&\lbrack X+\xi, Y+\eta\rbrack_F +i_{\lbrack X, Y\rbrack}B+i_Yi_X dB-di_Yi_X B
\nn\\
&=&e^B\bigg(\lbrack X+\xi, Y+\eta\rbrack_F\bigg)+i_Yi_X dB-di_Yi_X B.
\nn\\
\eea
We cannot only use $dB=0$ to show the automorphism. This indicates that we lose the $O(D, D)$ structure if we insist on including the one-form gauge field in this theory. The modification of the $O(D, D)$ structure can be seen from the modification of the bracket structures which comes from the total derivative term. The $C$-bracket and the $F$-bracket give the same gauge transformation of the metric, antisymmetric background and field strength. The total derivative term should come from the one-form gague field.

The double objects are not necessary to discuss the existence of the $O(D, D)$ structure. With the information of the diffemorphism and one-form gauge transformation, and field contents, the information is enough to discuss the existence of the $O(D, D)$ structure. After explaining this issue, we construct the action from symmetry point of view. We construct the action in two parts. The first part is the DBI part that we already mentioned. The other part is the background fields without involving the gauge field for the two derivative terms. Now we discuss how to formulate this part. This is interesting for the form of the action with the two derivative terms is uniquely determined based on the $O(D, D)$ structure, $\mathbb{Z}_2$ symmetry, gauge symmetry with the strong constraints.  We first discuss the $\mathbb{Z}_2$ symmetry
\bea
B_{MN}\rightarrow -B_{MN},
\qquad
\tilde{\partial}^M\rightarrow -\tilde{\partial}^M.
\eea
This implies
\bea
{\cal E}_{MN}\rightarrow{\cal E}_{NM}.
\eea
We rewrite $\tilde{\partial}^M\rightarrow -\tilde{\partial}^M$ as
\be
\partial_A\rightarrow  Z\, \partial_A \,.
\ee
The off-diagonal matrices of the ${\cal H}^{AB}$
change sign under the transformation $B_{MN} \rightarrow - B_{MN}$. This shows
\be
{\cal H}^{AB}  \to  Z {\cal H}^{AB} Z\,, \qquad
{\cal H}_{AB}  \to  Z {\cal H}_{AB} Z \,.
\ee
Then we construct the action from the gauge symmetry (with the strong constraints) from all possible $O(D, D)$ elements ($\partial_A$, ${\cal H}^{AB}$, ${\cal H}_{AB}$ and $d$) up to a boundary term. The action is
\bea
\label{acg}
S_2 &=& \int dx \ d\tilde x  \
   e^{-2d}\Big(\frac{1}{8}{\cal H}^{AB}\partial_{A}{\cal H}^{CD}
  \partial_{B}{\cal H}_{CD}
\nn\\
&&
-\frac{1}{2}
  \,{\cal H}^{AB}\partial_{B}{\cal H}^{CD}\partial_{D}
  {\cal H}_{AC}
\nn\\
  &&-2\partial_{A}d\partial_{B}{\cal H}^{AB}+4{\cal H}^{AB}\,\partial_{A}d
  \partial_{B}d \Big).
 \eea
 This action is uniquely determined from the above requirement.
The action of the DBI part is 
\bea
S_1 &=& \int dx \ d\tilde x  \ e^{-d}\bigg(-\det( t-F )\bigg)^{\frac{1}{2}}\frac{1}{(-\det \frac{t+t^t}{2})^{\frac{1}{4}}}.
\nn\\
\eea
Here, we use $e^{-d}$ for the DBI action because this term shows the manifest equivalence of the commutative and non-commutative gauge theories. This is equivalent to saying that this term is invariant by exchanging closed and open string parameters. This scalar dilaton term also has the manifest equivalence on the Buscher rule.
A total action of a space-filling brane is
\bea
S&=&S_1+\alpha S_2
\nn\\
&=&\int dx \ d\tilde x  \ e^{-d}\Big[\bigg(-\det( t-F )\bigg)^{\frac{1}{2}}\frac{1}{(-\det \frac{t+t^t}{2})^{\frac{1}{4}}}\Big]
\nn\\
&&+\alpha e^{-2d}\Big[\bigg(\frac{1}{8}{\cal H}^{AB}\partial_{A}{\cal H}^{CD}
  \partial_{B}{\cal H}_{CD}
\nn\\
&&-\frac{1}{2}
  \,{\cal H}^{AB}\partial_{B}{\cal H}^{CD}\partial_{D}
  {\cal H}_{AC}
\nn\\
  &&-2\partial_{A}d\partial_{B}{\cal H}^{AB}+4{\cal H}^{AB}\,\partial_{A}d
  \partial_{B}d\bigg) \Big],
\eea
where $\alpha$ is an arbitrary constant.
Using $\tilde{\partial}^M$=0, we obtain
\bea
&&\int dx\ \sqrt{-\det g}\bigg[e^{-\phi}\bigg( -\det(g+B-F)\bigg)^{\frac{1}{2}}
\nn\\
&&\times\bigg( -\det g\bigg)^{-\frac{1}{2}}
\nn\\
&&
+\alpha e^{-2\phi}\bigg(R+4(\partial\phi)^2-\frac{1}{12}H^2\bigg)\bigg],
\nn\\
\eea
where $R$ is the Ricci scalar and $H=dB$ is the three form field strength. This action is a low-energy effective theory from the combination of the closed and open strings. A consistent double sigma model with non-constant background fields should give this space-filling action from the one-loop $\beta$ function. Now we use the symmetry point of view to construct this action which is exactly consistent with \cite{Callan:1986bc}. The coefficient $\alpha$ can be determined from the one-loop $\beta$ function.
We define the generalized scalar curvature in this double field theory based on the symmetry and equation of motion of the scalar dilaton as
\bea
{\cal R}&\equiv&
\frac{1}{2}\bigg(-\det( t-F )\bigg)^{\frac{1}{2}}\frac{1}{(-\det \frac{t+t^t}{2})^{\frac{1}{4}}}
\nn\\
 &&+\alpha\bigg(4{\cal H}^{AB}\partial_A\partial_B d-\partial_A\partial_B{\cal H}^{AB}
\nn\\
&&-4{\cal H}^{AB}\partial_A d\partial_B d+4\partial_A{\cal H}^{AB}\partial_B d
\nn\\
&&+\frac{1}{8}{\cal H}^{AB}\partial_A{\cal H}^{CD}\partial_B{\cal H}_{CD}
\nn\\
&&-\frac{1}{2}{\cal H}^{AB}\partial_A{\cal H}^{CD}\partial_C{\cal H}_{BD}\bigg).
\eea
By using $\tilde{\partial}^M$=0 without considering the DBI term, the gauge transformation of the generalized scalar curvature satisfies
\bea
\delta_{\xi}{\cal R}=\xi^A\partial_A{\cal R}.
\eea
The DBI term will break this symmetry.
This shows the difference between the closed and open string theories. We can reinterpret the $F$-bracket by
\bea
\lbrack\xi_1, \xi_2\rbrack_F^A=\lbrack\xi_1, \xi_2\rbrack_C^A-\frac{1}{2}\partial^A\bigg(\xi_{2D}Z^D{}_{E}\xi_1^E\bigg).
\nn\\
\eea
The difference between the $F$-bracket and $C$-bracket is $-\frac{1}{2}\partial^A\bigg(\xi_{2D}Z^D{}_{E}\xi_1^E\bigg)$. If we use $\tilde{\partial}^M$=0, the difference is an exact one form. From this construction, we can easily understand that the candidate of the suitable gauge transformation should be different from the Courant bracket by an exact one form. This kind of the deformation from the one-form gauge fields should be expected. This also explains why we {\it never} use the twisted Courant bracket ($\lbrack X+\xi, Y+\eta\rbrack_{\mbox{twist}}\equiv\lbrack X+\xi, Y+\eta\rbrack_{\mbox{Cour}}+i_Yi_XH$) as the gauge transformation of the massless closed string theory. 

\section{Double Sigma Model}
\label{5}
We discuss the double sigma model \cite{Copland:2011wx} and start from 
\bea
\label{bulk}
S_{\mbox{bulk}}&=&-\frac{1}{2}\int d^2\sigma\bigg(-\partial_1X^A{\cal H}_{AB}\partial_1X^B
\nn\\
&&+\partial_1X^A\eta_{AB}\partial_0X^B\bigg).
\eea
The double sigma model (\ref{bulk}) gives the same equation of motion as in the ordinary sigma model for the flat worldsheet metric with $(-, +)$ signature on the bulk. An equation of motion of (\ref{bulk}) is
\bea
&&\partial_1\bigg({\cal H}_{AB}\partial_1X^B-\eta_{AB}\partial_0X^B\bigg)
\nn\\
&&=\frac{1}{2}\partial_1 X^B\partial_A{\cal H}_{BC}\partial_1X^C.
\nn\\
\eea
To show classical equivalence between the double and ordinary sigma models, we use $\tilde{\partial}^M$=0. Then we have
\bea
\label{con1}
{\cal H}^M{}_B\partial_1X^B-\eta^M{}_B\partial_0X^B=0
\eea
to remove half degrees of freedom. This is equivalent to
\bea
g^{-1}\partial_1\tilde{X}-g^{-1}B\partial_1 X-\partial_0 X=0.
\eea
For convenience, we rewrite this as
\bea
\partial_1\tilde{X}=B\partial_1 X+g\partial_0 X.
\eea
The gauge transformation of $X$ is governed by the generalized Lie derivative. The generalized Lie derivative is
\bea
\hat{{\cal L}}_{\xi} V^A=\xi^C\partial_C V^A+(\partial^A\xi_C-\partial_C\xi^A)V^C.
\nn\\
\eea
The gauge transformation of the background fields is
\bea
\delta g_{MN}&=&{\cal L}_{\epsilon} g_{MN},
\nn\\
\delta B_{MN}&=&\partial_M\Lambda_N-\partial_N\Lambda_M+{\cal L}_{\epsilon}B_{MN}
\nn\\
\eea
with $\tilde{\partial}^M$=0. The gauge parameters do not depend on the worldsheet coordinates. Therefore, we find (\ref{con1}) is covariant under the gauge transformation with $\tilde{\partial}^M$=0. This implies that we do not need to modify (\ref{con1}) when including the one-form gauge field.
We substitute (\ref{con1}) to the other equation of motion, then we obtain
\bea
\label{cl1}
&&\partial_1\bigg({\cal H}_{MB}\partial_1X^B-\eta_{MB}\partial_0X^B\bigg)
\nn\\
&=&\partial_1\bigg(Bg^{-1}\partial_1\tilde{X}+(g-Bg^{-1}B)\partial_1X-\partial_0\tilde{X}\bigg)_M
\nn\\
&=&\partial_1\bigg(Bg^{-1}\partial_1\tilde{X}+(g-Bg^{-1}B)\partial_1X\bigg)_M
\nn\\
&&-\partial_0(g\partial_0X+B\partial_1X)_M
\nn\\
&=&\partial_1(g\partial_1 X+B\partial_0 X)_M-\partial_0(g\partial_0X+B\partial_1X)_M,
\nn\\
\eea
\bea
\label{cl2}
&&\frac{1}{2}\partial_1 X^B\partial_M{\cal H}_{BC}\partial_1X^C
\nn\\
&=&\frac{1}{2}\partial_1\tilde{X}\partial_Mg^{-1}\partial_1\tilde{X}+\partial_1X\partial_M(Bg^{-1})\partial_1\tilde{X}
\nn\\
&&+\frac{1}{2}\partial_1X\partial_M(g-Bg^{-1}B)\partial_1X
\nn\\
&=&-\frac{1}{2}\partial_0X\partial_Mg\partial_0X+\frac{1}{2}\partial_1X\partial_Mg\partial_1X
\nn\\
&&+\partial_1X\partial_MB\partial_0X.
\eea
We combine (\ref{cl1}) and (\ref{cl2}) to find the same equation of motion as the equation of motion in the ordinary sigma model
\bea
&&\frac{1}{2}\int d^2\sigma\bigg(\partial_{\alpha}X^M g_{MN}\partial^{\alpha}X^N
\nn\\
&&-\epsilon^{\alpha\beta}\partial_{\alpha}X^MB_{MN}\partial_{\beta}X^N\bigg).
\eea
If we impose the Neumann boundary condition in the $\sigma^1$ direction, the boundary term
\bea
S_{\mbox{boundary}}=-\int d\sigma^0\ A_M\partial_0X^M
\eea 
 is necessary for the gauge invariance and boundary condition. This boundary term breaks the $O(D, D)$ structure, which is consistent with the previous understanding. Since the DBI theory on the constant background comes from fluctuation of the gauge field, we cannot write the DBI theory in terms of the $O(D, D)$ elements. However, the above double sigma model already has the classical equivalence. It should be interesting to compute the one-loop $\beta$ function to get the DBI theory from quantum fluctuation. The most important thing is that this double sigma model is computable.

\section{Conclusion}
\label{6}
We construct the double field theory of the DBI theory and the double sigma model of the open string. The construction of the DBI theory is based on the generalized geometry. It is interesting to understand the effect on the one-form gauge field. This is equivalent to comparing the $C$-bracket with the $F$-bracket. The $F$-bracket implies that the DBI action cannot be written down by the $O(D, D)$ elements. Based on the symmetry principles, we write down the suitable form for the action with the non-trivial flux. This action also reflects the difference of the T-duality between the closed and open strings. Even if we lose the manifest T-duality at the level of action, this does not imply that we cannot obtain lower dimensional theories. In the case of the massless closed string theory, we change variables from commutative to non-commutative descriptions. This change of variables are the manifest T-duality in closed string theory. For obtaining lower dimensional theories, we perform compactification. For the open string, we can use the same way to obtain lower dimensional theories. In the double sigma model, we find a boundary term or the gauge field that breaks the $O(D, D)$ structure. All of these give a consistent understanding with the $F$-bracket. We define the generalized scalar curvature based on the symmetry and equation of motion of the scalar dilaton. This generalized scalar curvature contains the DBI term and the low-energy massless closed string. Our construction of the double sigma model has a strong evidence on the classical equivalence. We only need to include the boundary term to get the Neumann boundary condition and gauge invariance without modifying other relations. This should be the minimum modification.

Double field theory with local gauge symmetries relies on the strong constraints. This is a famous drawback in the double formulation. But we should keep in mind that a formulation without local symmetries is also useful. The entanglement entropy \cite{Casini:2011kv} suffers from local gauge symmetries problem. In the low-energy effective theory of closed string, we cannot define the gauge invariant entanglement entropy. Double formulation for the low-energy effective closed string theory should avoid this problem. 

The study of the D-brane should inspire us to construct the M5-brane theory. The double field theory provides more constraints for us to construct the action. For the D-brane theory, we can determine this theory based on the symmetry point of view except for a relative coefficient. If this kind of symmetry point of view can be a principle in the brane theory, this should be interesting. A similar principle for the M5-brane theory should also be significant. 

From the study of the double sigma model of the closed string, we already understood the equivalence from the one-loop $\beta$ function, which is consistent at the level of quantum fluctuation. However, the classical equivalence does not imply that we should have quantum equivalence. This is a nontrivial consistent check on this theory and should show more interesting new physics to the double formulation.

\section*{Acknowledgement}
The author would like to thank Wu-Yen Chuang, Xing Huang, Pei-Wen Peggy Kao, Feng-Li Lin, Hisayoshi Muraki, Peter Schupp, Satoshi Watamura and Barton Zwiebach for their useful discussion. This work is supported in part by NTU (grant \#NTU-CDP-
102R7708), National Science Council (grant \#101-2112-M-002-027-MY3), CASTS (grant \#103R891003), Taiwan, R.O.C..

\end{document}